\documentclass[12pt]{article}
\usepackage{epsfig, latexsym}
\title{Generalized Hawking-Page phase transition}
\author{Parthasarathi 
Majumdar\footnote{parthasarathi.majumdar@saha.ac.in} \\ Saha Institute of
Nuclear Physics \\Kolkata 700 064, India.}
\begin{document}
\maketitle
\begin{abstract}

The issue of radiant spherical black holes being in stable thermal
equilibrium with their radiation bath is reconsidered. Using a simple
equilibrium statistical mechanical analysis incorporating Gaussian
thermal fluctuations in a canonical ensemble of isolated horizons, the
heat capacity is shown to diverge at a critical value of the classical
mass of the isolated horizon, given (in Planckian units) by the {\it
microcanonical} entropy calculated using Loop Quantum Gravity. The
analysis reproduces  the Hawking-Page phase transition discerned for
anti-de Sitter black holes and generalizes it in the sense that
nowhere is any classical metric made use of.

\end{abstract}

\section{Introduction}

A fairly complete understanding of the entropy of isolated,
macroscopic, generic four dimensional general relativistic black holes
has been reached \cite{pm1} within the framework of the isolated
horizon \cite{abf1} and its quantization in Loop Quantum Gravity
\cite{abck}. Not only has the Bekenstein-Hawking area law been
reproduced (albeit with a fit of the Barbero-Immirzi parameter)
\cite{abck}, and an
infinite series of finite, unambiguous {\it quantum} corrections has
been found \cite{km1},\cite{km2}, with the leading correction,
logarithmic in the horizon area, showing a degree of universality. 

Radiant black holes, in contrast, are not directly accessible by these
methods. Indeed, isolated horizons, by virtue of their isolation, {\it
cannot} radiate. However, a heuristic approach employing equilibrium
statistical
mechanics including Gaussian thermal fluctuations has been applied
\cite{dmb}, \cite{cm1}-\cite{cm5}, \cite{pm2} to unravel the canonical
(and {\it grand} canonical) entropy of radiating spherical black
holes. Building on these results, this paper attempts to associate a 
phase
structure with such black holes, and also to reveal the
relevance of the structure of quantum spacetime in arriving
at this phase structure. Since no classical metrics are used anywhere
in the analysis, in contrast to the incipient work of Hawking and Page
\cite{hp} (on the thermal stability of spherical anti-de Sitter
black holes), this work is to be thought of as a generalization of that
work. The analysis makes crucial use of the existence of 
quantum corrections to the Bekenstein-Hawking area law for
microcanonical entropy found within the  Loop Quantum Gravity
perspective \cite{km1}. 

The paper is organized as follows: in section
II the quantum Hamiltonian constraint is shown to imply that  
canonical partition function can be expressed entirely in terms of a sum
over states characterizing the event horizon (treated as a boundary of
the black hole spacetime). In the continuum limit, the boundary
partition function is expressed as an integral over horizon areas. The
saddle point computation of the boundary partition function (with an {\it
isolated} horizon of fixed area as the saddle point) is shown to lead
to the same universal criterion for thermal stability obtained earlier
by us \cite{cm1} somewhat differently. In the next section, the phase 
structure of this
thermodynamic system is investigated. It is shown that the transition
from the stable to the unstable phase is a first order phase transition
where the canonical heat capacity diverges. This phase transition
can be thought of as a generalization of the first order phase 
transition between thermally stable and unstable black holes
discerned by Hawking and Page \cite{hp} within a semiclassical
analysis restricted to asymptotically anti-de Sitter spacetimes. In
the concluding section, we point out certain {\it exact} properties of 
the boundary partition function, which has
the potential to lead to a phase structure {\it beyond the `mean field
theory' usually implied by the saddle point approximation}.    

\section{A Universal Stability Criterion}

To begin with, we adhere to the standard definition of a classical 
black hole
spacetime as ${\cal B} \equiv {\cal M} - J^-({\cal I}^+)$ where ${\cal
  M}$ is the entire spacetime and the $J^-$ is the part of spacetime
in the chronological past of future null infinity ${\cal I}^+$ for
asymptotically flat spacetimes. For asymptotically simple spacetimes
like (anti)-de Sitter, the corresponding infinity is taken. Consider a 
canonical ensemble of such black hole spacetimes
coupled to a radiation bath at
a temperature $\beta^{-1}$. The canonical partition function can be
written as 
\begin{eqnarray}
Z(\beta) ~=~ Tr \exp -\beta {\hat H}
\label{partf}
\end{eqnarray}
where, the total Hamiltonian operator ${\hat H}$ can be decomposed as 
${\hat H}  = {\hat H}_v + {\hat
H}_b$. The trace is over composite (spacetime + matter) states 
which can be expanded over a generic basis set
\begin{eqnarray}
|\Psi \rangle ~=~ \sum_{v,b} c_{vb} |\psi_v\rangle |\chi_b \rangle,
\label{stat}
\end{eqnarray}
where $|\psi_v\rangle$ represent composite states involving spin
network-like states of bulk
space within a Canonical Quantum Gravity scenario (like Loop Quantum
Gravity for instance, \cite{alrev}) and bulk quantum matter states, while
$|\chi_b\rangle$  represent composite gravity-matter states associated
with boundaries of spacetime. We {\it assume} that bulk states
$|\psi_v\rangle$  are annihilated by the bulk Hamiltonian 
\begin{eqnarray}
{\hat H}_v |\psi_v\rangle =0~ \label{annh}
\end{eqnarray}
as dictated by temporal diffeomorphism invariance. The trace then
reduces to
\begin{eqnarray}
Z &=& \sum_{b} \left( \sum_v |c_{vb}|^2 || ~|\psi_v\rangle~||^2
\right) \langle \chi_b|\exp - \beta {\hat H}_{bdy} |\chi_b
\rangle \nonumber \\
&=& Tr_{bdy} \exp -\beta {\hat H}_{bdy} \nonumber \\
& \equiv & Z_{bdy} ~,
\label{bparf}
\end{eqnarray}
where, we have assumed that the composite matter-space bulk states are
{\it normalizable} \footnote{It is not clear if this is a valid 
assumption in the case of Loop Quantum Gravity, see \cite{alrev}.} so 
that $\sum_v |c_{vb}|^2 ||~|\psi_v \rangle~||^2 = \xi_b^2 \in [0, \infty)$. 

One concludes that bulk states have 
no direct contribution to the thermodynamics of gravitating systems with
boundary; the latter is completely determined by 
states (degrees of freedom) on the boundary. One can think of this as
a more general notion of {\it holography} than propounded earlier
\cite{gth}. The earlier idea essentially means that the boundary
degrees of freedom contains the entire information of the bulk. But
in diffeomorphism invariant systems, the bulk Hamiltonian must
annihilate physical states, hence the entire information {\it is} what is
encoded in the boundary degrees of freedom. Is there then no role of
the bulk degrees of freedom ? The bulk states contribute as {\it
sources} to the boundary dynamics. However, once this dynamics is
determined by solving the boundary quantum equations of motion, this
is sufficient to determine the thermodynamics of the system. 

Thus, in the particular case that the boundary in question is a black
hole horizon, this is perhaps plausible evidence that its canonical
entropy must be a function of some attribute of the horizon, like its
area, although our approach does not actually prove that. To
proceed further, we make one additional assumption : the
eigenvalues of the Hamiltonian (which we call `mass') associated with 
the boundary are 
a function of the eigenvalues of the area operator associated with
the horizon, as determined within Loop Quantum Gravity. If
$g({ M}_n)$ is the degeneracy of the
boundary state with eigenvalue ${ M}_n$ , the boundary partition
function can be written as 
\begin{eqnarray}
Z(\beta) ~=~ \sum_n g({ M}(A_n)) \exp -\beta { M}(A_n)~, 
\label{bparf2}
\end{eqnarray}
where, we focus on the {\it large area} eigenvalues of the boundary: $
A_n = 4\pi \sqrt{3} \gamma l_P^2 \cdot n$ for $n \gg 1$. As in
\cite{cm2}, this enables us to write the partition function as an
integral over the energy
\begin{eqnarray}
Z(\beta) ~=~ \int d{ M} \exp \left [ S_{micro}({ M} ) - \beta
{ M} - \log \left| {d{ M} \over dn} \right| \right] ~, 
\label{bparf3}
\end{eqnarray}
we have used the standard relation between degeneracy of states and
the microcanonical entropy, and the last term in the exponent is
merely a Jacobian factor when one changes integration variables from
$n$ to ${ M}$. Since $dA(n) /dn = const.$, the partition function
can be recast in terms of an integral over $A$
\begin{eqnarray}
Z(\beta) ~=~\int_0^{\infty} dA~ \exp \left[ S_{micro}(A) - \beta {M}(A) \right] ~\label{areq}
\end{eqnarray}

The integral in (\ref{areq}) is performed using
the saddle point approximation \cite{cm2}, with the saddle point
chosen to be the area $A_{IH}$ the equilibrium horizon which we assume
to be an {\it isolated} horizon. Including the effect of Gaussian 
thermal fluctuations of the area
around the equilibrium configuration, one obtains
\begin{eqnarray}
Z_{IH} \simeq \exp \left [ S_{micro}(A_{IH}) -\beta M_{IH}(A_{IH}) \right ]  \cdot \left[ {\pi \over
S_{micro~AA}(M_{IH})} \right]^{1/2}~, 
\label{sad}
\end{eqnarray}
where, the last square-root factor is the contribution of the Gaussian
fluctuations around $A_{IH}$, with the subscripts implying derivatives
with respect to area. Using the statistical mechanical
definition of the canonical entropy, one gets \cite{cm2}, \cite{cm5}
\begin{eqnarray}
S_{canon}~=~S_{micro} -\frac12 \log \Delta ~,
\label{scan}
\end{eqnarray}
where, 
\begin{eqnarray}
\Delta \equiv {k \over M_A(A)}~ \left[M_{AA}(A) S_A(A) - M_A(A)
S_{AA}(A) \right]~,
\label{delta}
\end{eqnarray}
where, $k$ is a positive number independent of the area and we have
dropped the subscript $IH$ from $M$ and $A$ and $micro$ from $S$ which
now represents the microcanonical entropy. The primes indicate
derivatives with respect to the argument. It is clear that a {\it
sufficient} condition for the isolated horizon being a point of stable
thermal equilibrium is that 
\begin{eqnarray}
\Delta(A) ~>~0 \label{stab}
\end{eqnarray}
which is precisely the condition found in \cite{cm5}. 

The heat capacity can also be calculated and one
obtains \cite{cm5}, using the microcanonical definition of temperature
\begin{eqnarray}
C~= ~ S_A^2(A)~\Delta^{-1}(A).
\label{spht}
\end{eqnarray}
This equation now leads to the same condition (\ref{stab}) as the {\it
necessary} condition for thermal stability. Thus, spherical black
holes are thermally stable {\it iff} (\ref{stab}) holds. It is not
difficult to integrate this differential inequality with respect to
area and obtain, with a
few dimensional constants chosen appropriately for convenience, 
\begin{eqnarray}
M(A) ~>~ S(A)~. \label{condi}
\end{eqnarray}
If we put back dimensional fundamental constants, this condition can
be expressed as expressed as 
\begin{eqnarray}
M(A)~>~ \left({\hbar c \over G k_B^2} \right)^{1/2} ~S(A) ~,\label{prev}
\end{eqnarray}
an inequality obtained earlier \cite{cm5} somewhat differently. 

\section{A generalized Hawking-Page phase transition} 

Clearly, {\bf $C \rightarrow \infty$ as
$\Delta(A) \rightarrow 0$} thereby signifying a {\it first order phase
transition}. It is easy to see that this phase transition happens for
a critical area $A_c$ of the horizon, for which
\begin{eqnarray}
M''(A_c)S'(A_c) ~=~ M'(A_c) S''(A_c)
\label{crit}
\end{eqnarray}
The canonical entropy, and hence the free energy have a discontinuity at
this critical value of the area. This in turn can be used to define a
critical mass $M_c \equiv M(A_c)$. The variation of
the heat capacity as a function of the equilibrium mass is depicted in
Fig. 1. 
\begin{center}
\epsfxsize=90mm
\epsfbox{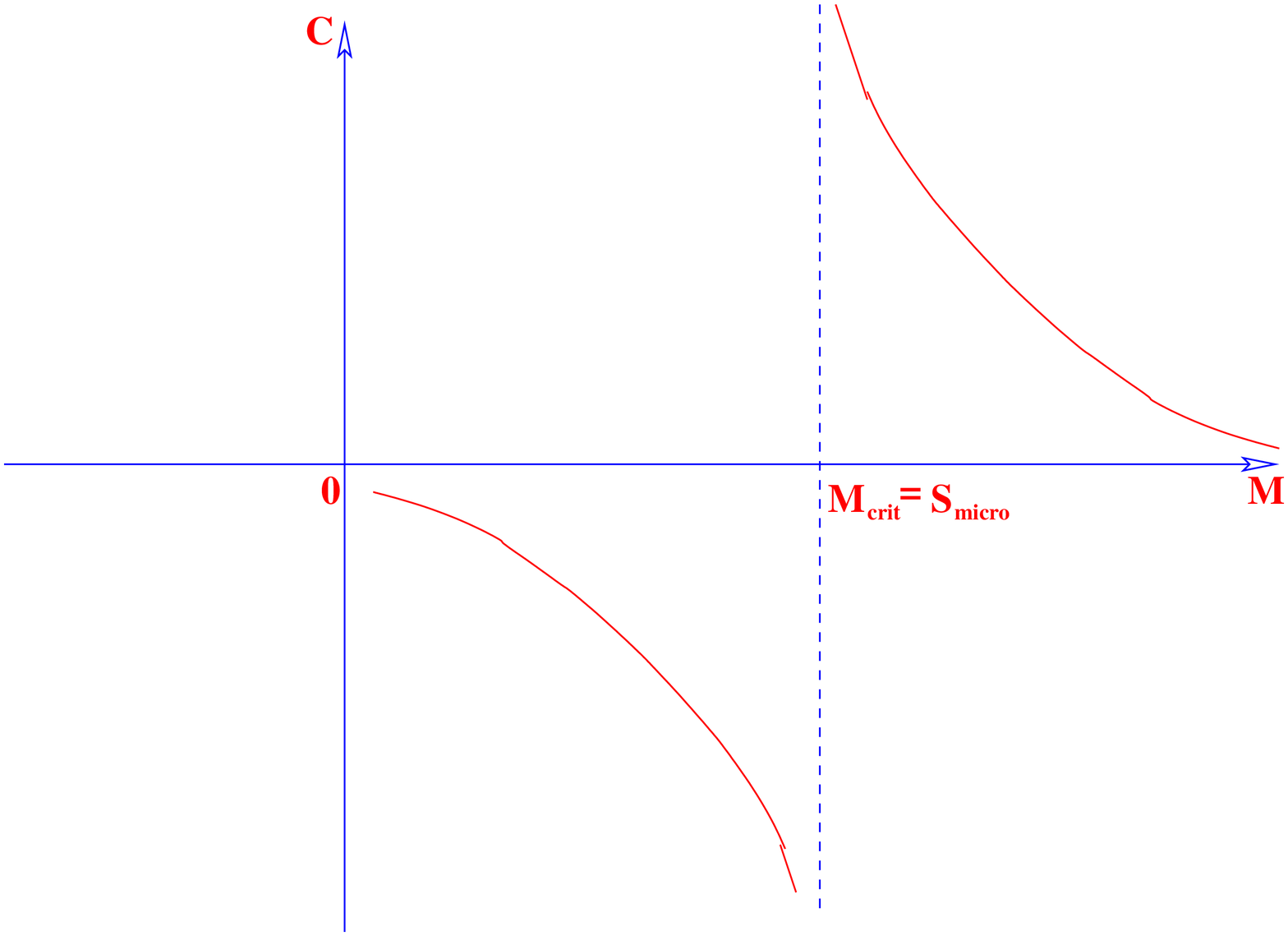}

{Fig. 1 {\em {Heat capacity as a function of equilibrium mass}}}
\end{center}

The phase transition thus occurs exactly when the mass of the
equilibrium configuration, as a function of the classical area is
{\it identical} (in Planckian units) with the microcanonical entropy
calculated from Loop Quantum Gravity and includes the infinite series
of corrections over and above the area term \cite{km1}. In fact, the
existence of
at least the leading logarithmic correction with a negative
coefficient turns out to be particularly important, for it ensures that  
$S_{AA}(A) >0$ which is required for the existence of a phase boundary at
the critical mass (see Fig. 2). 
\begin{center}
\epsfxsize=60mm
\epsfbox{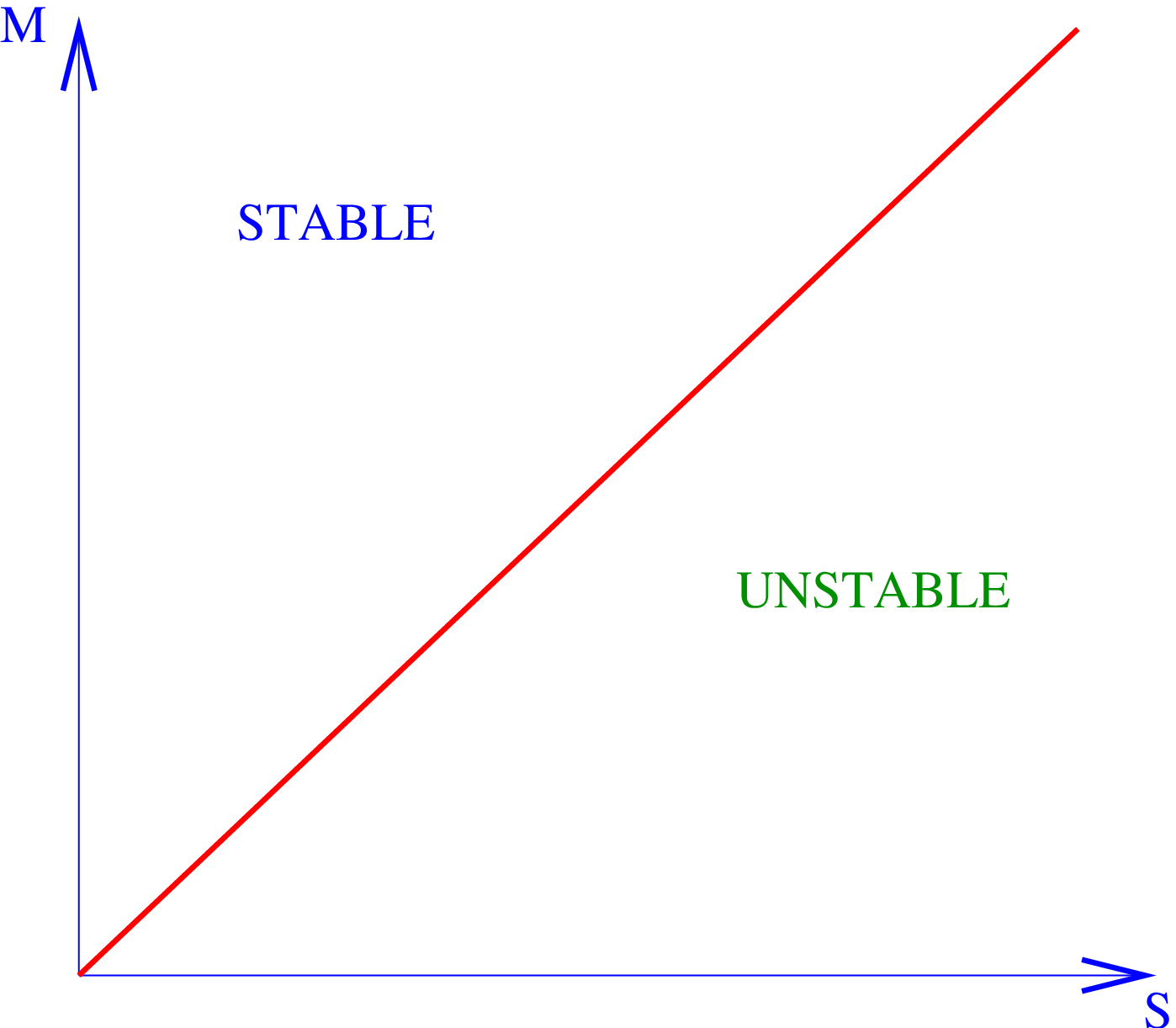}

{Fig. 2 {\em {Phase diagram showing stable and unstable thermal
equilibrium phases}}}
\end{center}

This phase
boundary separates the phases corresponding to a black hole being in
stable or unstable thermal equilibrium with respect to the radiation
bath, given by the condition $M(A) > S(A)$  or $M(A) < S(A)$, as
already discussed in \cite{cm5}. In terms of the effective
microcanonical temperature, this corresponds respectively to $T > 1$
or $T<1$. 

In the foregoing analysis, no assumption has been made regarding the
functional dependence of the equilibrium isolated horizon mass $M(A)$;
such a dependence of course originates from the behaviour of the
classical metric corresponding to the black hole spacetime. E.g., for
the ordinary Schwarzschild black hole, it is known that $M(A) \sim
A^{1/2}$ and therefore it does not satisfy our criterion above for
thermal stability, as is well-known to be the case. The same is true
for the Reisnner-Nordstrom black hole as can be easily seen by
extracting the mass as a function of the horizon area for very large
areas. We are of course assuming that the radiation from the black
hole is composed of neutral, rather than charged, particles. In the
latter event, the analysis will have to be done within a {\it grand}
canonical ensemble \cite{pm2} with perhaps similar end results. Thus
asymptotically flat spherical black holes tend to have only an
unstable thermal phase in which they radiate or accrete
interminably. We are excluding here the possibility of the existence
of yet another phase considered in \cite{hp} where thermal
stability of pure radiation against collapse into black holes has been
considered. Our analysis is valid for spacetimes {\it with an inner
boundary} and does not address the dynamical issue of how that
boundary came about.  

Going over now to non-asymptotically flat geometries, the most
important one is of course the adS variety, considered in detail in
ref. \cite{hp}. From the adS-Schwarzschild metric, it is easy to
read-off the mass-area relation
\begin{eqnarray}
M(A) ~=~\frac12 \left( {A \over 4\pi }\right)^{1/2}~\left(1 + {A \over
4\pi l^2} \right)~, 
\label{ads}
\end{eqnarray}
where, $l \equiv (-\Lambda)^{-1/2}$ upto a numerical factor. The
existence of two phases is obvious. If $\Lambda$ is large, i.e., if
the horizon area $A> 4\pi l^2$ substantially, to leading order in this
area (in Planck units) $M(A) \sim A^{3/2}$. Then, as is clear from
Fig. 2, one obtains stable thermal equilibrium. On the other hand, if
$\Lambda$ is smaller, such that $A \sim 4\pi l^2$, then $M(A) \sim
A^{1/2}$ once again. One is then clearly
in an unstable thermal equilibrium phase, since this is precisely the
mass-area relation for an ordinary (asymptotically flat) Schwarzschild 
geometry, which  corresponds to negative specific heat. Thus. the two 
phases
precisely correspond to the two black hole phases discerned in
ref. \cite{hp} for such a spacetime. Yet, nowhere in our analysis have
we made any reference to classical spacetimes.

What about the point of transition ? Our analysis, depicted in Fig. 1
would lead to the conclusion that the transition is first order at a
critical mass equal to the microcanonical entropy as a function of the
area. It is useful to compare this with the Hawking-Page transition
from a stable adS-Schwarzschild black hole (for $A >> 4\pi l^2$) to an
unstable one (for $A \sim 4\pi l^2$). One can determine the precise
point of transition in terms of $A$ by referring to \cite{dmb}; using
eq. (28) of ref. \cite{dmb}, the heat capacity for an
adS-Schwarzschild in four dimensions is given by 
\begin{eqnarray}
C~=~{A \over 4 l_P^2}~ \left[ {{6A \over 4\pi l^2}~+~1 \over {6A \over 
4\pi l^2} ~-~ 1} \right] ~. \label{adshc}
\end{eqnarray}
The critical area is clearly $A_c=\frac{2\pi}{3} l^2$. This is
certainly consistent with that obtained using our formula (\ref{crit})
and the adS-Schwarzschild mass-area relation (\ref{ads}). Similar
conclusions emerge for the generic ads-Reissner-Nordstrom solution as
well, of course restricted to the canonical ensemble for a neutral
radiation bath. Once again, no recourse is anywhere taken to classical
spacetime metrics in our analysis.

\section{Beyond the saddle-point approximation}

The saddle point approximation used in the analysis so far is similar
to the approximation used in the incipient work of Hawking and Page to
evaluate their partition function expressed as a functional integral
over Euclidean metrics. The conclusions
we have drawn therefore may be compared with theirs  based
upon their choice of the saddle point in the Euclidean functional 
integral to
be the classical metric itself, together with appropriate boundary
contributions. The improvement in our approach is that we have made no
explicit reference to the classical metric, and hence have what may be
thought of as a {\it generalization} of that earlier work. However,
the use of the saddle point approximation. which may be thought of as
`a mean field theory' in the sense of many body theory, is often not
considered reliable, since during a real phase transition,
`collective' effects may actually be quite large
and thus lead to a rather different phase behaviour. If so, both the
original Hawking-Page phase transition and our generalization of it
will be subject to non-trivial corrections beyond the saddle point
approximation. We have not made any substantive progress towards
delineating the precise nature of these corrections within our
approach. However, from a consideration of the starting formula
(\ref{areq}) for the boundary partition function expressed as an
integral over the horizon area, some facets of a phase structure
appear to emerge without a detailed analysis. 

Since both the mass and the microcanonical entropy are positive and
monotonically increasing functions of the horizon area, it is obvious 
that the partition function in (\ref{areq}) is finite provided 
\begin{eqnarray}
\beta M(A) ~>~ S(A) ~
\end{eqnarray}
or, if
\begin{eqnarray} 
\beta ~>~ \beta_{c0} ~\equiv~ {S(A) \over M(A)}  \label{partfi}
\end{eqnarray}
where the microcanonical entropy and mass are to be evaluated {\it
at equilibrium} where one can assume that they refer to an isolated
horizon for which both quantities are well-defined.  
 
Similarly, the thermally averaged mass, defined to be the equilibrium
mass, is finite provided 
\begin{eqnarray}
\beta ~>~ \beta_{c1} ~\equiv~ { {S(A) + \log M(A)} \over M(A)} 
\label{massfi}
\end{eqnarray}
i.e., at a higher critical inverse temperature of the heat
bath. Finally, the heat capacity is finite provided
\begin{eqnarray}
\beta ~>~ \beta_{c2} ~\equiv~ { {S(A) + 2\log M(A)} \over M(A)} 
\label{sphtfi}
\end{eqnarray}
an even higher inverse temperature. In this range, the heat capacity
can be expressed as 
\begin{eqnarray}
C~=~\langle M^2 \rangle_{\beta} ~-~\langle M \rangle^2_{\beta}
~\equiv~ \Delta M^2~, 
\label{hcpos}
\end{eqnarray}
where, $\Delta M^2$ is the mean squared fluctuation of mass, an obviously
positive quantity, corresponding to a stable thermal equilibrium
phase. However, for $\beta \leq \beta_{c2}$, the heat capacity appears
to {\it diverge}, in general, thus strongly suggestive of a first
order phase transition.  

The consistency of these general behaviour patterns with those
obtained in the earlier sections through the saddle point
approximation is not obvious. We hope to report on this and other
aspects of these exact formulae elsewhere. 

\vglue .3cm

\noindent {\it Acknowledgment}

We thank T. Harmark for an incisive question which
led to this investigation and the Particle Theory group at the Niels
Bohr Institute, Copenhagen for hospitality during which the present
work crystallized. We also thank Ashok Chatterjee, Ayan Chatterjee,
Saurya Das, Amit Ghosh and Parthasarathi Mitra for helpful discussions.

\end{document}